\documentclass[12pt,preprint]{aastex}
\newcommand\beq{\begin{equation}}
\newcommand\eeq{\end{equation}}
\newcommand\beqar{\begin{eqnarray}}
\newcommand\eeqar{\end{eqnarray}}

\begin{document}

\title{MAGNETICALLY-CONTROLED SPASMODIC ACCRETION DURING STAR FORMATION: 
II. RESULTS} 

\author{Konstantinos Tassis \& Telemachos Ch. Mouschovias}

\affil{Departments of Physics and Astronomy \\
University of Illinois at Urbana-Champaign, 1002 W. Green Street, Urbana, IL 61801}

\begin{abstract}

The problem of the late accretion phase
of the evolution of an axisymmetric, isothermal magnetic disk surrounding a forming
star has been formulated in a companion paper. The ``central sink approximation'' 
is used to circumvent the problem of describing the evolution inside the 
opaque central region for
densities greater than $10^{11}$ $\rm{cm}^{-3}$ and radii smaller than a few AUs.
Only the electrons are 
assumed to be attached to the magnetic field lines, and 
the effects of both negatively and positively charged grains are accounted
for. 

After a mass of $0.1$ $M_{\odot}$ accumulates in the central cell
(forming star), a series of magnetically
driven outflows and associated outward propagating shocks form
in a quasi-periodic fashion. As a result, mass accretion onto the protostar
occurs in magnetically controlled bursts. We refer to this process as
spasmodic accretion. The shocks propagate
outward with supermagnetosonic speeds. The period of dissipation
and revival of the outflow decreases in time, as the mass
accumulated in the central sink increases. We evaluate the
contribution of ambipolar diffusion to the
resolution of the magnetic flux problem of star formation during the
accretion phase, and we find
it to be very significant although not sufficient to resolve the entire
problem yet. Ohmic dissipation is completely negligible in the disk during
this phase of the evolution. The protostellar disk is found to be
stable against interchange-like instabilities, despite the fact that
the mass-to-flux ratio has temporary local maxima.

\end{abstract}


\section{Introduction}

In an accompanying paper (Tassis \& Mouschovias 2004, hereafter Paper I)
we formulated the problem of the formation and evolution of a nonrotating 
protostellar fragment in a manner that allows accurate modeling of the physics 
of the protostellar disk without the complication introduced by radiative transfer 
when the central protostar becomes opaque. In the present paper we follow the 
evolution until the mass of the central protostar grows to $1$ $M_{\odot}$.
The importance and relevance of this problem 
as well as previous work on the subject were summarized in \S 1 of Paper I. 
The evolution of the system is described by the six-fluid MHD equations, 
in which neutral molecules, ions, electrons, neutral grains, positively and 
negatively charged grains are treated as distinct but interacting fluids. 
We use the ``central sink'' method to study the structure and evolution of
the isothermal disk surrounding the forming protostar.
The effects of the mass and magnetic
flux accumulating in the central, opaque protostar on the isothermal disk are 
accounted for.

We include important physics ignored by
previous calculations: (1) the decoupling of the 
ions from the magnetic field lines, which occurs at densities above 
$10^8 \, {\rm cm ^{-3}}$ (Desch \& Mouschovias 2001) and which leaves 
the magnetic field frozen only in the much less massive and much more
tenuous electron fluid; (2) the chemical and dynamical effects
of the positively-charged grains, 
the abundance of which becomes significant in the same density regime. 
Phenomena that arise due to this new physics are discovered, such 
as the formation and dissipation of
a series of shocks in a quasi-periodic fashion. 

In \S 2 we summarize the most important results of the contraction 
phase prior to the introduction of the central sink. The evolution at later times,
in the presence of a central sink, is described in \S 3. Specifically, we discuss:
the evolution 
of the system until the first electron outflow occurs (\S 3.1); the establishment of 
a quasi-periodic magnetic cycle (\S 3.2) and the properties of one typical
such cycle (\S 3.3) ; the time evolution of the mass, magnetic flux, and
mass-to-flux ratio in the central protostar (\S 3.4); the properties of the 
shock in the neutral fluid (\S 3.5); and the evolution of the period 
of the magnetic cycle (\S 3.6). Finally, we investigate the stability of
the supercritical core against a magnetic interchange instability (\S 3.7). 
The most important conclusions are summarized
in \S 4.

\section{Contraction Prior to the Central Sink Approximation}

In order to obtain the proper initial conditions for the calculation of the collapsing
cloud after the introduction of the central sink, we 
follow the ambipolar-diffusion--induced formation and contraction of a
protostellar core in the absence of a central sink up to a central
neutral density in excess of $10^{11}$ ${\rm cm^{-3}}$. The initial state is
an exact equilibrium state in the absence of ambipolar diffusion, as in
Ciolek \& Mouschovias (1993).
We reproduce the results presented by Desch and Mouschovias (2001), 
who studied the formation of protostars up
to a central density $\gtrsim 10^{12}$ ${\rm cm^{-3}}$ using a
modified ZEUS-2D code. This attests to the accuracy of the two very
different numerical codes. We
summarize the main features of the evolution, emphasizing those that
are different from the corresponding
results of Ciolek and K\"{o}nigl (1998), because of our more accurate
physical assumptions (i.e., flux-freezing in the electrons instead of the ions, and
inclusion of positive grains) and more appropriate numerical treatment of both 
the $r$-component of the magnetic field and the boundary conditions at the surface
of the central sink, at $r=R{\rm inner}=7.31$ AU.

Starting from a reference state at temperature $T=10$ K, a uniform
magnetic field $B_{\rm ref} = 35.26$
${\rm \mu G}$, and a central mass-to-flux ratio $\mu_{\rm c,0} =
0.256$, in units of the critical value for collapse,
we calculate the initial equilibrium state. The
model cloud in equilibrium has a total mass $M = 96.6$ $M_{\odot}$ and central
number density $n_{\rm n,c0} = 2700$ ${\rm cm^{-3}}$. It is thermally
supercritical (the Bonnor-Ebert critical mass is only $M_{BE} = 3.5 / /
M_{\odot}$; see Bonnor 1956; Ebert 1955, 1957)
but magnetically subcritical. At time $t = 0$ we start the simulation by
``turning on'' ambipolar diffusion. 

%%%%%%%%%%%%%%%%%%%% FIGURE %%%%%%%%%%%%%%%%%%%%%%%

\begin{figure}
\plotone{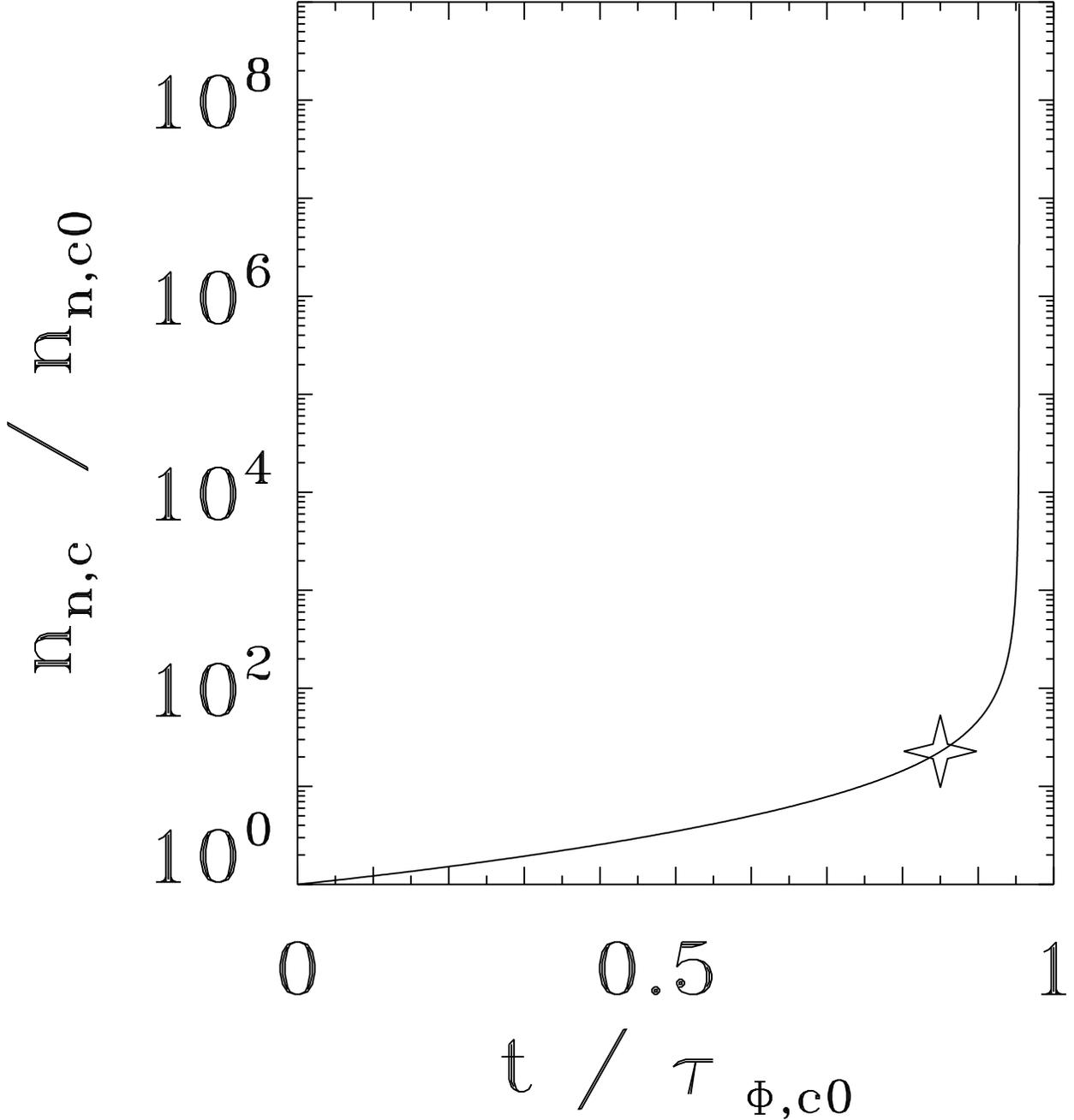}
\caption{\label{fig_ncent}Evolution of central number density of neutrals normalized to
   its initial value $n_{\rm n,c0} = 2.7 \times 10^3$ ${\rm cm^{-3}}$ as a
   function of time (normalized to the
  initial central flux-loss timescale, $\tau_{\rm \Phi,c0} = 1.88
   \times 10^7$ yr) {\em before} the introduction
  of the central sink. The ``star'' on the curve marks the instant at
   which the central mass-to-flux ratio becomes equal to its critical
   value for collapse.}
\end{figure}

Figure \ref{fig_ncent} shows the evolution of the central neutral
density, $n_{\rm n,c}$, normalized
to its initial equilibrium value $n_{\rm n,c0}$, as a function of
time, normalized to the initial central flux-loss timescale,
$\tau_{\Phi,c0} \equiv [\Phi_B / (d\Phi_B/dt)]_{c0}$. The two
distinct phases in the evolution of the model cloud are 
evident. Initially, during the ambipolar-diffusion--controlled phase, the
central density increases slowly, at the initial central
flux-loss time, by a factor 
$\simeq 30$ in the first $1.598 \times 10^7$ ${\rm yr}$. By then the central
mass-to-flux ratio has exceeded the critical value for collapse and
the evolution changes dramatically
after this time. The now magnetically (and thermally) supercritical
core contracts dynamically (but is not free-falling) and
the central density increases (from $\approx 7 \times 10^4$ ${\rm
  cm^{-3}}$) rapidly by seven orders of magnitude in the next
$2 \times 10^5$ ${\rm yr}$. 
  
%%%%%%%%%%%%%%%%%%%% FIGURE %%%%%%%%%%%%%%%%%%%%%%%
\begin{figure}
\plotone{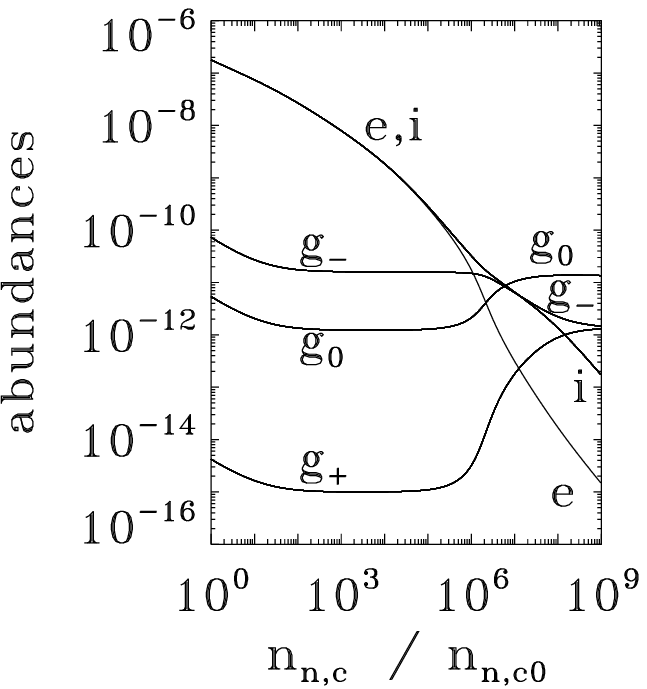}
\caption{\label{fig_chem}
Central abundances (electrons e, ions i, negative grains ${\rm g_-}$, positive grains
${\rm g_+}$, and neutral grains ${\rm g_0}$) relative to the total density of neutrals versus
the central neutral density $n_{\rm n,c}$, normalized as in Fig. 1.}
\end{figure}

The central relative abundances of all species, $\chi_{\rm s} \equiv
n_{\rm s}/n_{n}$ (where $s = {\rm i}$, e, ${\rm g_+}$, ${\rm g_-}$, ${\rm g_0}$)
are shown
in Figure \ref{fig_chem} as functions of the neutral density.  For low densities,
corresponding to the cloud envelopes and the early stages of the
evolution of the central region,
the atomic and molecular ions are slightly more abundant than the
electrons and much more abundant than any grain species. The grains are primarily negatively
charged because of more efficient electron (than ion) attachment onto
neutral grains.
During the early, ambipolar-diffusion--controlled phase of
the evolution, the fractional abundances of all grain
species decrease, because they are attached to magnetic field lines
and are ``left behind'' as the neutrals, under the action of their
self-gravity, contract through all charged species. Thus the
total central dust-to-gas mass ratio decreases initially. For
densities higher than $10^{10}$ ${\rm cm^{-3}}$, the 
grains become the dominant charge carriers. Now the grains are primarily
neutral and, for even higher densities, the positive grains are almost as abundant as
the negative grains. Since the electrons are depleted (by attachment
onto the neutral grains) faster than the ions, the ions become considerably more
abundant than the electrons.

%%%%%%%%%%%%%%%%%%%% FIGURE %%%%%%%%%%%%%%%%%%%%%%%

\begin{figure}
\plotone{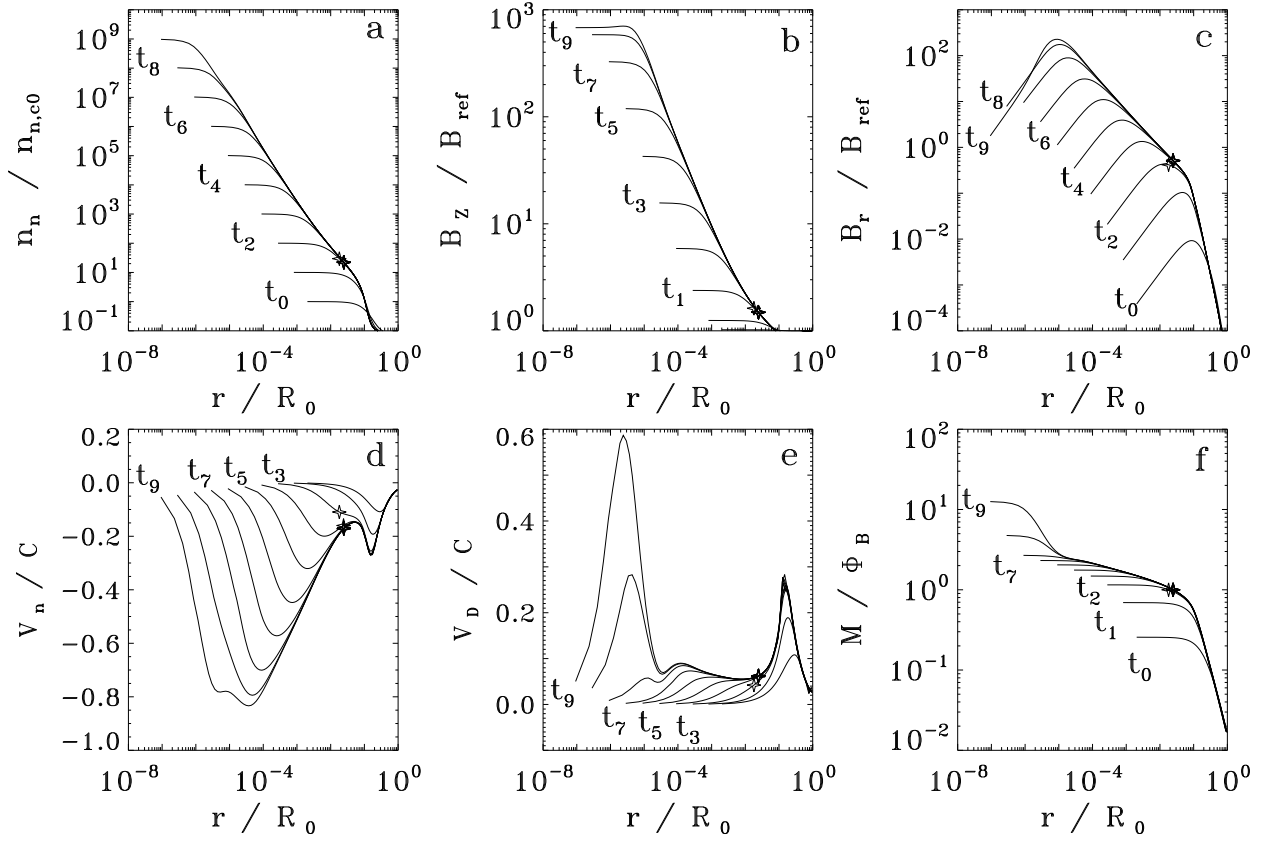}
\caption{\label{fig_precsa_profiles}
   Ambipolar-diffusion-induced evolution of model cloud. Spatial profiles
   are shown at ten different times before the introduction
   of the central sink, namely at $t$ = 0, 13.66, 16.99, 17.43,
  17.49, 17.5044, 17.5072, 17.5079, 17.50812, and 17.50816 Myr. The radial
  coordinate is normalized to $R_0 = 4.23$ pc. (a) Neutral density
   $n_{\rm n}$, normalized to its initial central equilibrium value
   $n_{\rm n,c0}$ $(= 2.7 \times 10^3$ ${\rm cm^{-3}}$); 
(b) $z$-component of the magnetic field $B_z$ normalized to the
  reference-state magnetic field, $B_{\rm ref}$ $(=35.27$ ${\rm \mu G}$);
(c) $r$-component of the magnetic field $B_r$ normalized to the
  reference-state magnetic field, $B_{\rm ref}$;
(d) radial velocity of the neutrals $v_{\rm n}$ normalized to the isothermal sound
  speed, $C$ $(=0.188$ km ${\rm s^{-1}})$;
(e) radial drift velocity $v_{\rm D}$ between electrons and neutrals normalized to
  the sound speed in the neutrals, $C$;
(f) mass-to-flux ratio $M/\Phi_{\rm B}$, normalized to the critical value for a thin disk 
  $1/(2\pi G^{1/2})$. } 
\end{figure}

Figures \ref{fig_precsa_profiles}a - \ref{fig_precsa_profiles}f 
show physical quantities as functions of the radial coordinate
$r$ (normalized to the initial cloud radius $R_0 = 4.23$ ${\rm pc}$) at ten
different 
times, $t_j$ $(j = 0,1,...,9)$. These times $t_j$ are chosen such that the
central density has increased by a factor $10^j$ $(j = 0,1,...,9)$ relative to
its initial equilibrium value. An asterisk on a curve marks the radius of the
magnetically supercritical core at that time; i.e., the total mass-to-flux
ratio within this radius is equal to the critical value. 

Figure \ref{fig_precsa_profiles}a 
exhibits the number density of the neutral particles (normalized to
its initial central value) as a function of radius at ten different
times (see caption for values). The density is almost
uniform in the innermost part of the core, where the thermal-pressure forces
dominate and smooth out all density fluctuations. At late times,
the density follows a nearly power-law dependence on the radius
outside the central region. The mean (logarithmic) slope in this
region is 1.7. The envelope
of the cloud where most of the mass resides hardly changes during the
evolution, remaining magnetically supported. The mass of the
supercritical core at the last time output, when $n_{n,c}/n_{n,c0} =
10^{9}$, is $7.7$ $M_{\odot}$, and an amount of mass equal to $1$
$M_{\odot}$ is contained within the radius $r/R_0 = 5 \times
10^{-3}$, i.e., inside $r=4.36 \times 10^3$ AU.

%%%%%%%%%%%%%%%%%%%% FIGURE %%%%%%%%%%%%%%%%%%%%%%%
\begin{figure}
\plotone{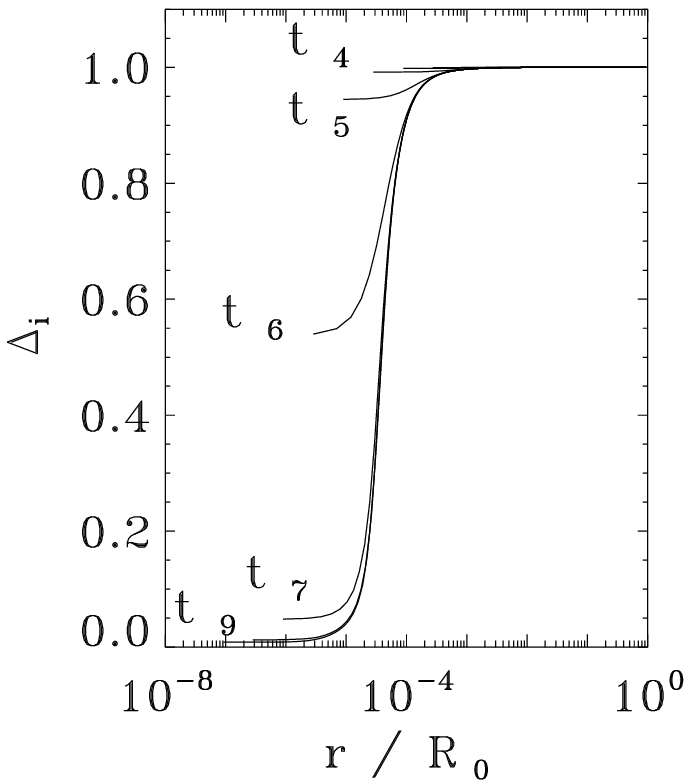}
\caption{\label{fig_deltai_precsa}
The ion attachment parameter, $\Delta_{\rm i}$, as a function of $r/R_0$
at the same ten times as in
Figure \ref{fig_precsa_profiles}. $\Delta_{\rm i}=1$ means perfect
attachment to, and $\Delta_{\rm i}=0$ complete detachment (of the ions)
from the field lines.}
\end{figure}

The spatial structure of the $z$-component of the magnetic field is
displayed in Figure \ref{fig_precsa_profiles}b, at the same ten times
as in Figure \ref{fig_precsa_profiles}a. It is qualitatively similar
to the density distribution, with a nearly uniform (but more extended) region near the
origin and an almost power-law region in the rest of the collapsing
core, characterized by a mean (logarithmic) slope of $0.7$. During the
dynamical stage of contraction, $B_{\rm c}
\propto n_{\rm c}^{0.47}$ (see review by Mouschovias 1987), so that each factor of
10 increase in $n_{\rm n,c}$ raises $B_{z,{\rm c}}$ by a factor of about 3.
This behavior breaks down above a central density of about $10^{10}$
${\rm cm^{-3}}$. Thereafter, the collapse of the matter in the
innermost region of the supercritical core does not much alter the
magnetic field structure because of re-awakening of ambipolar-diffusion.

The radial component of the
magnetic field, $B_r/B_{\rm ref}$, just above $z=Z$, is shown as a
function of radius in Figure \ref{fig_precsa_profiles}c, at the same
ten times as in Figure \ref{fig_precsa_profiles}a . Comparison to
Figure \ref{fig_precsa_profiles}b
reveals that $B_r<B_z$ at all times and at all radii.

Figure \ref{fig_precsa_profiles}d similarly
shows the radial infall velocity of the neutrals, $v_{{\rm n},r}$,
(normalized to the isothermal sound speed) as a function of $r/R_0$. Once a
magnetically supercritical core has formed, matter inside it contracts much more
rapidly than matter in the envelope. The infall velocity remains subsonic
everywhere throughout the run.

The drift velocity
$v_{\rm D} \equiv v_{{\rm e},r} - v_{{\rm n},r}$ (normalized to the
isothermal sound speed) is shown in Figure \ref{fig_precsa_profiles}e as a
function of $r/R_0$. Since electrons are attached to the magnetic
field lines, their velocity
represents that of the field lines as well.  Prior to the formation of the
supercritical core ($t < t_2$), ambipolar diffusion controls the
infall of the neutrals throughout the cloud. At $r/R_0 = 0.15$ both the
neutral infall speed (\ref{fig_precsa_profiles}d) and the drift speed
are maximum, with $v_{\rm D} \backsimeq -v_{{\rm n},r}$. These two
speeds are equal and opposite because the magnetic field lines are
essentially ``held in place'' (i.e., $|v_{{\rm e},r}| \ll |v_{{\rm
    n},r}|$) as the neutrals contract through them. Ambipolar
diffusion continues to control the infall of the neutrals in the
envelope at all times, while the smaller drift speed inside the
supercritical core at times $t_7 \geq t \geq t_2$ indicates the
dragging of the field lines inward with the neutrals. The increase of the drift velocity at
the innermost part of the supercritical core for densities above $10^{10}$ ${\rm
  cm^{-3}}$ ($t > t_7$) reveals that the neutrals are contracting significantly
faster than the field lines (rejuvenation of ambipolar diffusion) leading to the lack of increase of $B$
with increasing $n_{\rm n}$, as shown in Figure \ref{fig_precsa_profiles}b.

Figure \ref{fig_precsa_profiles}f shows the mass-to-flux ratio
$M/\Phi_{\rm B}$, 
normalized to the critical value,
$1/(2 \pi G^{1/2})$, as a function of $r/R_0$, at the same times as in
Figure \ref{fig_precsa_profiles}a. Initially the cloud
has a subcritical value at the center. The mass-to-flux ratio increases
during the ambipolar-diffusion--controlled phase of the evolution. A
supercritical core (marked by an asterisk) forms after 
the second curve. During the dynamical phase of
contraction the mass-to-flux ratio approaches an asymptotic value 2 - 3,
but after the density at the center exceeds $10^{10}$ ${\rm cm^{-3}}$ the
mass-to-flux ratio increases again, indicating the re-awakening of 
ambipolar diffusion, as found by Desch \& Mouschovias (2001).

Finally, the ion attachment parameter 
\beq
\Delta_{\rm i} \equiv \frac{v_{\rm i} - v_{\rm n}}{v_{\rm e} - v_{\rm n}}
\eeq
is shown as a function of $r/R_0$ in
Figure \ref{fig_deltai_precsa}, at the same times as in Figure \ref{fig_precsa_profiles}a. 
This parameter is a measure of how 
well the ions are coupled to the magnetic field. If $\Delta_{\rm i} = 1$, the ions
move with the electrons and the magnetic field, while if  $\Delta_{\rm} = 0$, the
ions are completely detached from the magnetic field and fall in with the
neutrals. Inside the region of radius $r \approx 2 \times 10^{-5} R_0
= 17.5$ AU, the ions begin to detach from the field lines at densities as low as
$10^8$ ${\rm cm^{-3}}$ (curve labeled $t_5$) and are almost completely
detached by a density
$10^{10}$ ${\rm cm^{-3}}$. This causes the re-awakening of ambipolar
diffusion, witnessed in 
Figures \ref{fig_precsa_profiles}b, \ref{fig_precsa_profiles}c,
\ref{fig_precsa_profiles}e and \ref{fig_precsa_profiles}f. The assumption of flux-freezing in the ion
fluid (used by Ciolek \& Konigl 1998) is therefore invalid for densities greater
than $10^8$ ${\rm cm^{-3}}$. 

\section{The CSA Calculation}

Now we introduce the central sink at radius $R_{\rm inner} = 7.31$ ${\rm
AU}$. The density at this position is $6.5 \times 10^{10}$ ${\rm cm^{-3}}$ and
the mass initially contained in the central sink is $M = 8 \times
10^{-4}$ ${\rm M_{\odot}}$, 
which is negligible compared to the mass of the cloud or the supercritical 
core ($M_{\rm core} = 7.6$ ${\rm M_{\odot}}$). We let the previous 
calculation go to
higher central densities ($\approx  
10^{12}$ ${\rm cm^{-3}}$) in order for the density profile of
Figure \ref{fig_precsa_profiles}a to
establish an almost power-low shape at all densities of interest, and the
central thermal lengthscale to become much smaller than the radius of the central
sink. For this part of the calculation we use a stationary grid with
160 zones.

\subsection{Evolution to Resurrection of Ambipolar Diffusion and \\
 First Electron Outflow}

\begin{figure}
\plotone{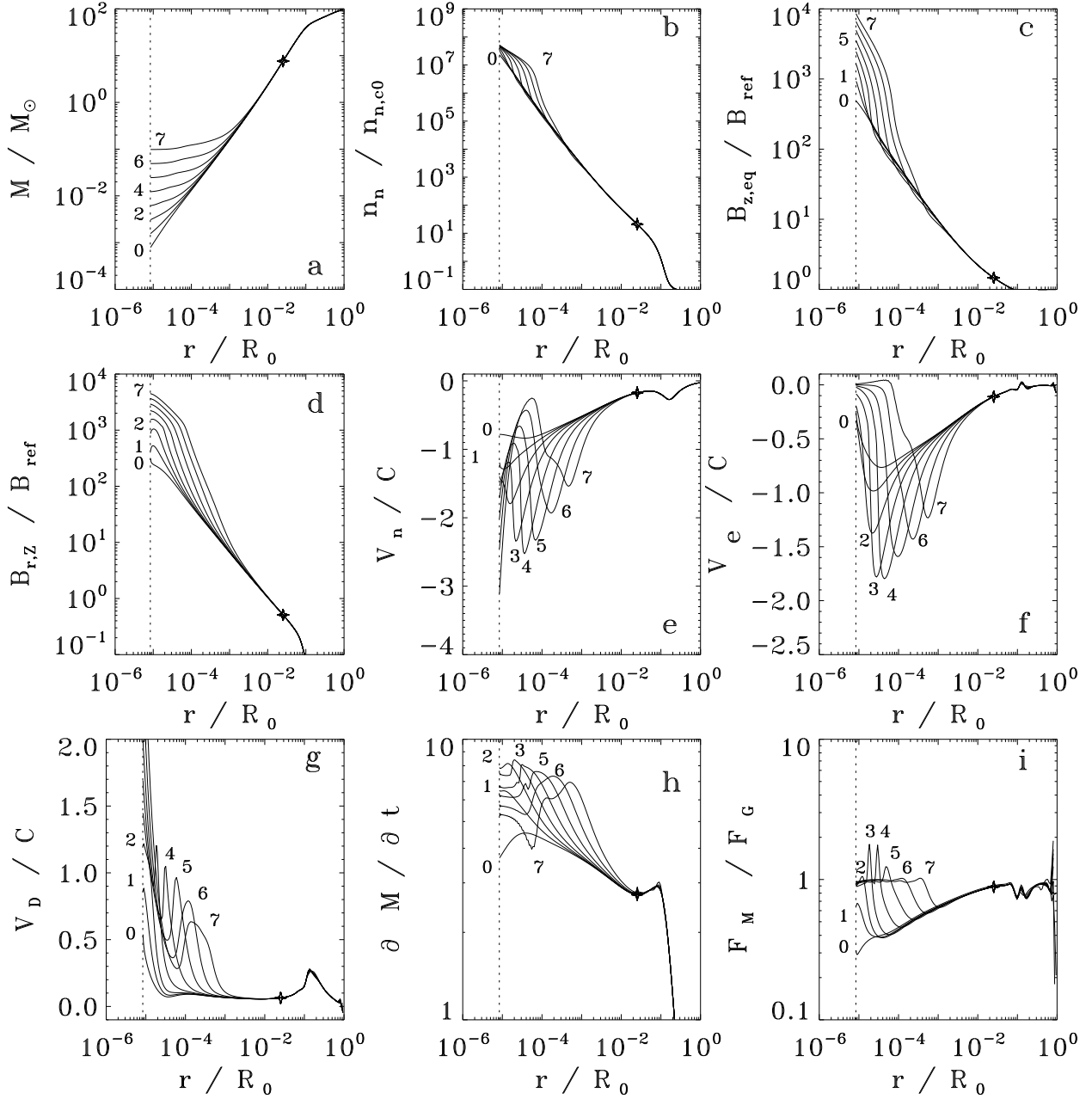}
\caption{\label{csafig1} 
   Spatial profiles of physical quantities at eight different times ($\Delta t$ = 0, 125, 316, 677, 1472,
   3213, 6983 and 15169 yr), after the central sink is introduced. The radial coordinate is normalized to 
   $R_0 = 4.23 pc$
   and the inner boundary (dashed line) is at $7.31$ AU. (a)
   Total mass in $M_{\odot}$; (b) Number density of neutrals normalized to its
   central value in the reference state, $n_{\rm n,c0} = 2.7
   \times 10^3$ ${\rm cm^{-3}}$; 
   (c) $B_z$ normalized to its value in the reference state, $B_{\rm ref} =
\,\, 35.26 {\rm \mu G} $; (d) $B_r$ normalized to $B_{\rm ref}$;
   (e) Radial velocity
   of neutrals, normalized to the isothermal sound speed, $C$; (f) Electron
   velocity, normalized to $C$; (g) Drift velocity of electrons with
   respect to the neutrals, normalized to $C$; (h) Mass accretion rate
   in ${\rm M_\odot/ Myr}$; (i) Ratio of magnetic and gravitational force.}
\end{figure}

Figure \ref{csafig1} shows various quantities as
functions of $r/R_0$, at eight different times 
($\Delta t$ = 0,  125, 316, 677,
1472, 3213, 6983 and 15170 yr). $\Delta t$ is the time elapsed since the
beginning of the CSA calculation. Each successive time marks the doubling of the mass
accumulated in the central sink. By the last time output shown in
Figure \ref{csafig1}a (curve 7), this mass is $0.1$
 ${\rm M_{\odot}}$. 
An asterisk on a curve locates, as in Figures 
\ref{fig_precsa_profiles}a-f, 
the instantaneous radius of the critical flux tube. The radius of the
supercritical core does not change from its initial value during the evolution. 

%%%%%%%%%%%%%%%%%%%% FIGURE %%%%%%%%%%%%%%%%%%%%%%%

As the evolution progresses, the neutral number density departs from
the established power-law of the pre-CSA evolution, and this deviation
propagates to larger and larger radii
(Fig. \ref{csafig1}b). A similar behavior is exhibited by the $z-$ and
$r-$ components of the magnetic field (Figs. \ref{csafig1}c and
\ref{csafig1}d respectively). In the region $ 5 \times 10^{-4} \lesssim r \lesssim 10^{-3} R_0$, $B_r$
becomes greater than $B_z$, implying
that magnetic-tension forces become more important than
magnetic-pressure forces in this region. However, in the cells adjacent to the
sink, $B_z$ is always larger than $B_r$. 

An important feature of the evolution of $B_z$
is the presence of a narrow radial region in which the field rises steeply
by almost an order of magnitude. The radial position of this {\em magnetic wall}
coincides with the position of a peak in the ratio of magnetic and
gravitational forces acting on the neutral fluid (Fig. \ref{csafig1}i). The
strong magnetic force decelerates the neutrals
(Fig. \ref{csafig1}e) and eventually causes a slight outflow (positive
velocity) in the electron fluid in the region $r \lesssim 7 \times 10^{-5}
R_0$ (last curve in Fig. \ref{csafig1}f). As we'll see below, this
magnetic wall forms, dissipates and reappears repeatedly during the
evolution of the model cloud.

The reason for the formation of the magnetic wall is revealed by the
time-evolution of the electron velocity profiles
(Fig. \ref{csafig1}f). Close to the sink, the resurrection of
ambipolar diffusion is manifested as a deceleration of the electron
fluid without a corresponding decrease of the velocity of the neutral
fluid. By $\Delta t = 3213 {\rm \, yr}$ (curve labeled by 5), the
electron infall speed
at the cell adjacent to the sink has decreased to zero, while at
larger radii the electrons are still falling in. This
causes a magnetic flux pileup, and further deceleration of the
electrons at greater and greater radii. Eventually, the magnetic force
becomes large enough to drive a slight outflow of the electrons in the
region $r \lesssim 7 \times 10^{-5} R_0 = 61$ AU  at time
$\Delta t = 15,184$ yr (see last curve in Fig. \ref{csafig1}f).

The response of the neutral fluid to the changes in the magnetic force
is shown in Figure \ref{csafig1}e, which exhibits the velocity of the
neutrals versus radius at the same times as in Figure \ref{csafig1}a.
By the time $\Delta t = 677$ yr (curve labeled by 3 in
Fig. \ref{csafig1}e) the
maximum infall speed has become supersonic (and
supermagnetosonic), and a shock has formed in the neutral fluid,
which propagates outward at subsequent times. At radii smaller than
the location of the shock front (i.e., behind the shock) the neutrals decelerate, 
and reach a minimum infalling speed at a radius
slightly smaller than the location of the maximum of the ratio $F_M/F_G$. At
yet smaller radii, the neutrals are
reaccelerated by the gravitational field of the mass already in the
central sink.

The resurrection of ambipolar diffusion is illustrated clearly in
Figure \ref{csafig1}g, which shows the drift velocity of the
electrons relative to the neutrals. Within the
supercritical core, the drift velocity develops a plateau at a low 
value, corresponding to the ineffectiveness of ambipolar diffusion
during a large part of the dynamical phase of contraction. However, a
peak develops in the
drift velocity at small radii at later times. This implies resurrection of ambipolar
diffusion because the ambipolar-diffusion timescale $\tau_{AD} \sim
r/v_D$ decreases locally. Inward of the peak of $v_D$ there is a local minimum, corresponding to
the local deceleration of the neutrals close to the position of the
magnetic force maximum. Closer to the boundary of the central sink, the
drift speed rises sharply and monotonically, because the electrons are
held back by the magnetic force while the neutrals get reaccelerated
by the mass already accumulated in the central sink. 

%%%%%%%%%%%%%%%%%%%% FIGURE %%%%%%%%%%%%%%%%%%%%%%%
\begin{figure}
\plotone{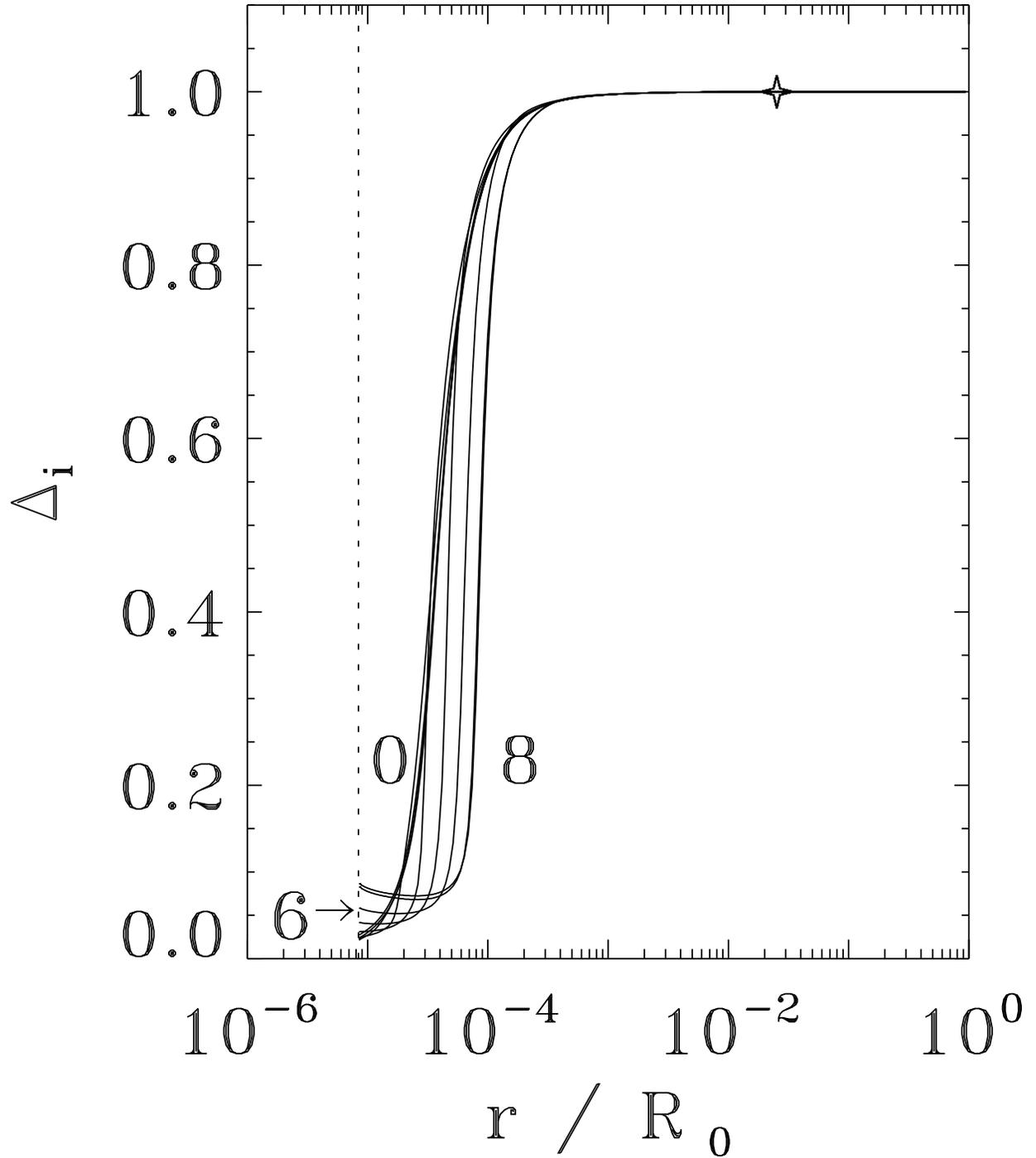}
\caption{\label{csafig_deltai} 
   Spatial profiles, after the central sink approximation, of the
   attachment parameter of the ions shown at the same times as in
   Fig. \ref{csafig1}.}
\end{figure}

The mass infall rate (Fig. \ref{csafig1}h) is almost a 
mirror image of the velocity of the neutral fluid. 
Initially it increases as $r$ decreases, but, after the formation of the shock in the neutrals,
it exhibits a steep drop at the position of the shock. This feature eventually
develops into a clear minimum, corresponding to the minimum infall
velocity of the neutral fluid. 

Finally, the ion attachment parameter is shown in Figure
\ref{csafig_deltai} as a function of $r/R_0$, at the same eight times
as in Figure \ref{csafig1}a. As
time progresses, the region of almost complete detachment (small values of $\Delta_{\rm
  i}$) of the ions from the field lines 
expands outward. This is a consequence of the significant increase of
the density at greater and greater radii (Fig. \ref{csafig1}b), deviating from the
power-law profile established in the pre-CSA calculation, without a corresponding
increase of $B_z$. An increase of the density leads to more
frequent collisions of ions with neutral particles causing them to detach
from the field lines, while an increase of the magnetic field leads to
better attachment of the ions.  By the time $\Delta t = 15,170 {\rm \,
  yr}$, at a radius of $r \approx 5 \times 10^{-5} R_0$ the
density has increased by almost two orders of magnitude, while $B_z$
has increased by only one order of magnitude. At the same time, at an
inner radius of $r \approx 10^{-5} R_0$ the density has increased by a factor of three
and $B_z$ has increased by more than an order of magnitude. Consequently 
at small radii, near the central-sink boundary, the ions become
somewhat better coupled to the field lines than they were at the
beginning of the run; $\Delta_i$ increases from $0.02$ to almost $0.1$.

\subsection{Quasi-Periodic Magnetic Cycle}

%%%%%%%%%%%%%%%%%%%% FIGURE %%%%%%%%%%%%%%%%%%%%%%%
\begin{figure}
\plotone{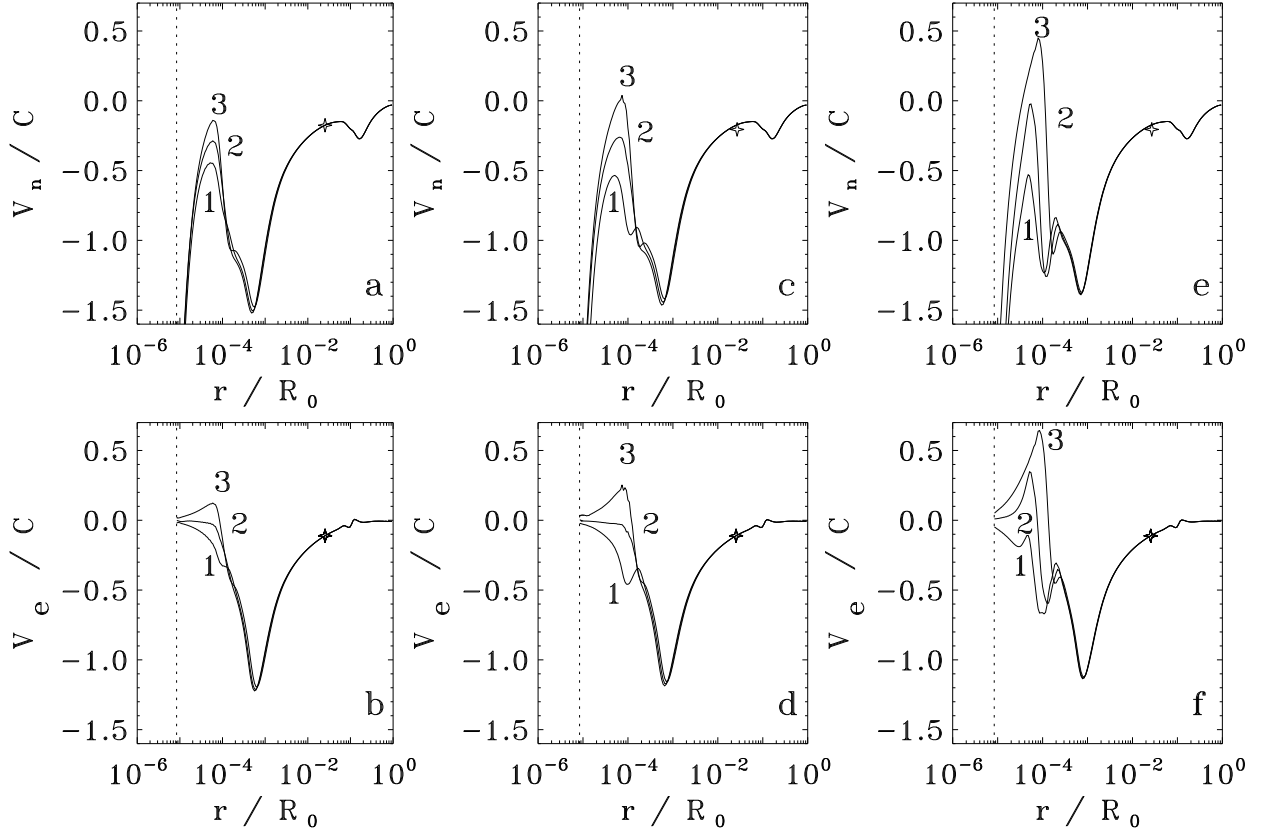}
\caption{\label{csa_show_periodicity} 
Profiles of the neutral velocity (upper panel) and the electron
velocity (lower panel) normalized to the isothermal sound speed in the
neutrals ($C = 0.188$ km ${\rm s^{-1}}$). The leftmost, middle, and 
rightmost frames correspond to the first, second and third magnetic 
cycles, respectively. Each curve in these plots corresponds to a
specific instant in each magnetic cycle, with the curve numbers (1-3) 
increasing from earlier to later times.}
\end{figure}

After the first electron (slow) outflow, the system enters a phase where the
magnetic flux transport, as well as the response of
the other quantities to the evolution of the magnetic flux, exhibit a 
quasi-periodic behavior.

To demonstrate this behavior, we have plotted in
Figure \ref{csa_show_periodicity} profiles of the velocities of the 
neutrals (upper panel) and the electrons (lower panel) for three
instances of each of the first three ``magnetic cycles'' after the
first electron (slow) outflow. The leftmost frames correspond to the first cycle, the
middle frames correspond to the second cycle, and the rightmost
frames to the third cycle. We focus our attention to the
region $r \lesssim 2 \times 10^{-4} R_0 = 174$ AU. 
The first (earliest) curve for each cycle
corresponds to infalling velocities at all radii for both neutrals and
electrons. Flux is being accumulated at small radii
during this phase of the cycle, since $|v_{\rm n}| \gg |v_{\rm e}|$. At the
instant corresponding to the second curve of each cycle, the
enhanced magnetic force due to the accumulated flux has reduced the electron velocity to
essentially zero and has significantly
decelerated the neutrals 
in the first two cycles while it has produced a mild
outflow in the third cycle. Finally, the third curve of each cycle
corresponds always to outflow in the electron fluid, and very small
infalling velocities (in the first cycle) or outflows (in the next
two cycles) in the neutrals. The outflow spreads the accumulated
flux to larger radii. As a result, gravity dominates again, and the
neutrals get reaccelerated inward, dragging with
them the electrons. The inflow of electrons recreates the
magnetic wall, and the cycle repeats itself. 

Other than the repetitive pattern of this behavior, it is also
important to note that the amplitude of the variation in the
velocities of the neutrals and electrons increases with each cycle. We
will discuss in detail the evolution of the amplitude and the period of the magnetic cycle after
we analyze the behavior of other physical quantities
during a typical cycle.

\subsection{Analysis of a Typical Magnetic Cycle: The Rise and Fall 
of a Magnetic Wall}

\begin{figure}
\plotone{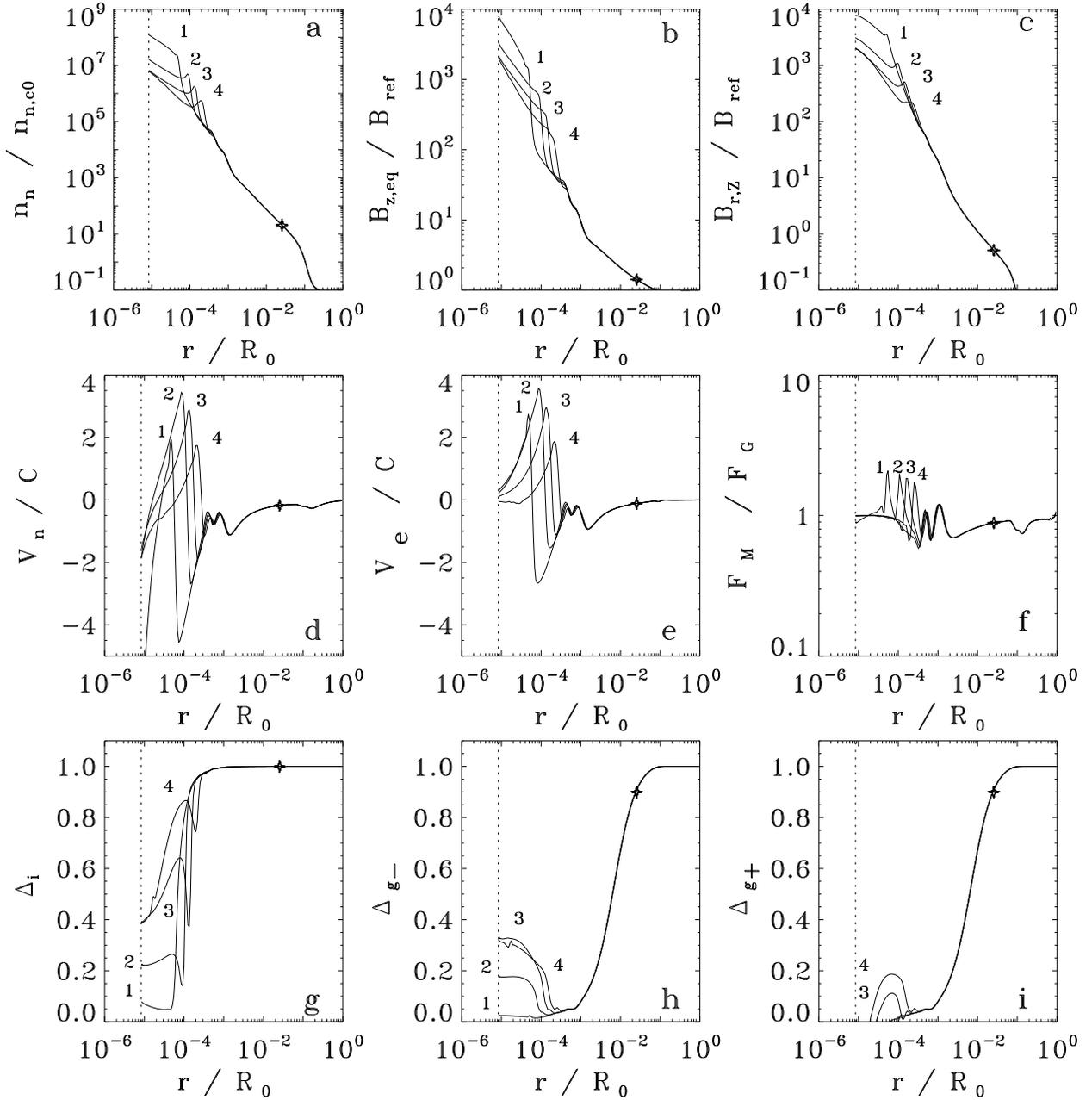}
\caption{\label{csafigna} 
   Spatial dependence of physical quantities during the first
   half-period of a magnetic cycle. Normalizations are as in Fig. 5.
  (a) Number density of neutrals;
   (b) $z-$component of the magnetic field $B_z$; (c) $r-$component of
   the magnetic field $B_r$;
   (d) Radial velocity  of neutrals $v_{\rm n}$;
   (e) Radial electron velocity $v_{\rm e}$;
   (f) Ratio of magnetic and gravitational forces $F_{\rm M}/F_{\rm G}$;
   (g) Attachment parameter of ions $\Delta_{\rm i}$;
   (h) Attachment parameter of negatively charged grains $\Delta_{\rm g-}$;
   (i) Attachment parameter of positively charged grains $\Delta_{\rm g+}$}
\end{figure}

\begin{figure}
\plotone{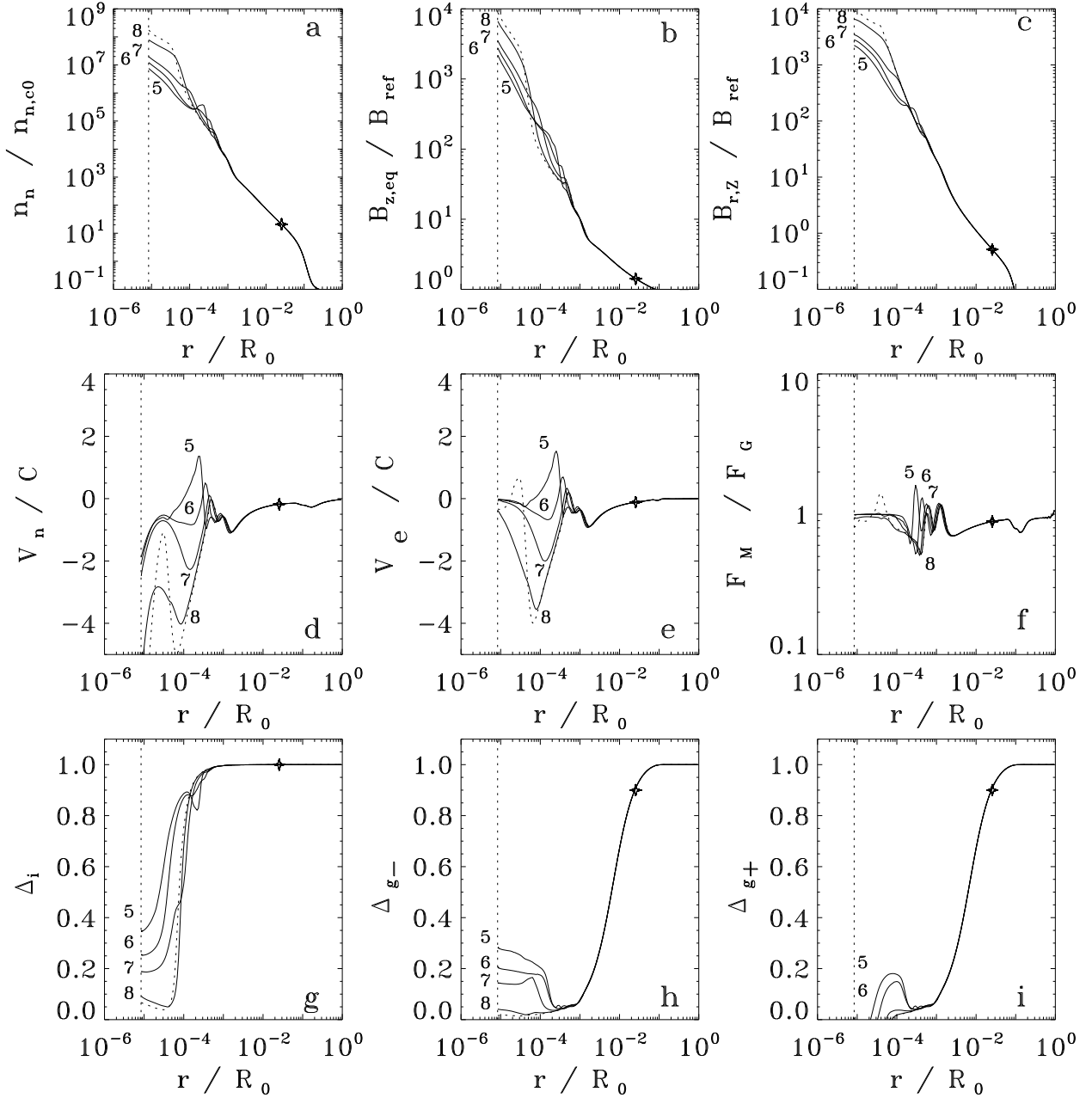}
\caption{\label{csafignb}
   Spatial profiles of the same quantities as in Fig. \ref{csafigna},
   but during the second half-period of a magnetic
   cycle.}
\end{figure}

Figures \ref{csafigna} and \ref{csafignb} show the evolution of the
radial profiles of the neutral
number density, magnetic field, radial velocity of the neutrals and
electrons, the ratio of magnetic and gravitational forces, and the
attachment parameters of the ions, the negative and positive grains during a
typical quasi-periodic cycle introduced above. For clarity,
the duration of a single cycle has been broken into two parts. During
the first part, shown in Figure \ref{csafigna}, at radii smaller than
the location of the maximum outflow velocity of the electrons
there is essentially no infall of magnetic flux, i.e., the velocity of the
electrons is either positive or almost zero. During the second part of the
cycle, shown in Figure \ref{csafignb}, the electrons are reaccelerated 
inward by the infalling neutrals (they acquire negative
velocities) at radii smaller than the position of 
the fastest outflow. The dashed curve in Figure \ref{csafignb} corresponds
to the first snapshot of the {\em next} cycle, showing the
re-establishment of outflow velocities close to the central sink. 

The number density of the neutral particles, as seen in 
Figure \ref{csafigna}a, decreases with time in the region where it has
deviated from its established global profile. This is so because some
matter moves outward, as seen in Figure \ref{csafigna}d (region of
positive $v_n$), while other matter falls into the central sink - in
the cells adjacent to the sink the neutrals still have negative
velocities. As time progresses, the number density of the
neutrals develops a local maximum at moderate radii, at the position
of the shock, where both the neutral outflow behind the shock as well
as the neutral infall ahead of the shock are responsible for a pileup
of matter. As the shock propagates outward, this local maximum
follows the shock, staying right behind it.

%%%%%%%%%%%%%%%%%%%% FIGURE %%%%%%%%%%%%%%%%%%%%%%%

The $z$-component of the magnetic field (Fig. \ref{csafigna}b) 
exhibits the sharp increase characteristic of the magnetic wall 
that drives the outflow. The position of this wall propagates
outward in time, while the
field strength at small radii decreases as the
electron outflow redistributes the magnetic flux outward and arrests
the accumulation of flux in this region. The $r-$component of the
magnetic field (Fig. \ref{csafigna}c),
which is a measure of the deformation of the field lines, also decreases
in time at small radii because of the local outflow of the electrons.
This tendency of the electron outflow to straighten out the field lines
at small radii is also responsible for the
development of a local maximum in the spatial profile of $B_r$ at the position of the
shock.

The most prominent features of the velocity profiles of the
neutrals and the electrons (Figs. \ref{csafigna}d and \ref{csafigna}e,
respectively) are of course the presence of the shock and the
outflows. The positions of the maximum outflow velocity and that of the
shock move outward, while the magnitude of the
outflow velocity eventually decreases as the outflow in the
electrons lowers the magnetic wall that drives it.
This physical picture is also supported by Figure \ref{csafigna}f, which
shows the ratio of magnetic and gravitational forces. The peak of
this ratio, identifying the position of the magnetic wall, does
indeed move outward while at the same time decreasing in magnitude. 

The attachment parameters of the ions and charged grains in
the region $r \lesssim 2 \times 10^{-4} R_0 = 175$ AU increase in time, in 
response to the significant decrease in
density (which is accompanied by only a moderate decrease of the
magnetic field). The effect is most pronounced for the ions and least
pronounced for the positively charged grains.

The direction of change of the quantities in the innermost radii is
reversed during the second part of the cycle, shown in
Figure \ref{csafignb}. The electron outflow in the first part of the
cycle creates a deficit of magnetic flux behind the position of the
maximum velocity outflow.  This deficit can be
seen in both the profiles of $B_z$ (Fig. \ref{csafignb}b), where a local minimum 
develops immediately behind the location of the magnetic
wall, and in the magnetic-to-gravitational force ratio
(Fig. \ref{csafignb}f) where a dip at values smaller than $1$ occurs
behind (at smaller radii than) the maximum. 

As a result, in the region of this magnetic deficit, the
gravitational force dominates and re-accelerates the
neutrals inward. Figure \ref{csafignb}d shows the progressively greater
infalling velocities acquired by the neutrals in the region behind the
outflow. In turn, the infalling neutrals drag with them the electrons,
and magnetic flux once again starts being transported inward behind the
shock. The magnetic wall will now disperse much faster than at earlier
times because
accumulated flux is transported both outward (due to the electron outflow
that continues) and inward (due to the
re-accelerated infall of the electrons). However, as the electrons acquire
greater infall velocities, a new accumulation of flux will occur near the original
position of the magnetic wall, and a new magnetic wall will form
there. Eventually, this magnetic wall drives a
new outflow, and a new magnetic cycle begins. The beginning of the new cycle
is marked by the dashed curve in the plots of
Figure \ref{csafignb}. Note that the electrons respond much faster than
the neutrals to the accumulation of magnetic flux. The neutrals
have to overcome their relatively large inertia, that tends to keep them falling
inward, before they can join the electron outflow. 

%%%%%%%%%%%%%%%%%%%% FIGURE %%%%%%%%%%%%%%%%%%%%%%%

The rest of the quantities in the second part of the cycle change in
response to the negative velocities acquired by the neutrals and the
electrons. The neutral number density at the inner radii increases in time
(Fig. \ref{csafignb}a) as new material starts flowing inward. 
In the same way, $B_z$ and $B_r$
(Figs. \ref{csafignb}b and \ref{csafignb}c, respectively) increase
at small radii in response to the inward transport of magnetic
flux by the electrons. Finally, the attachment parameters of ions and
charged grains decrease with time close to the central sink (Figs. \ref{csafignb}g,
\ref{csafignb}h and \ref{csafignb}i, respectively), in response to
the increasing number density of neutrals in the same region, which leads to more
frequent collisions and detachment of the charged species from the
field lines. 

%%%%%%%%%%%%%%%%%%%% FIGURE %%%%%%%%%%%%%%%%%%%%%%%
\begin{figure}
\plotone{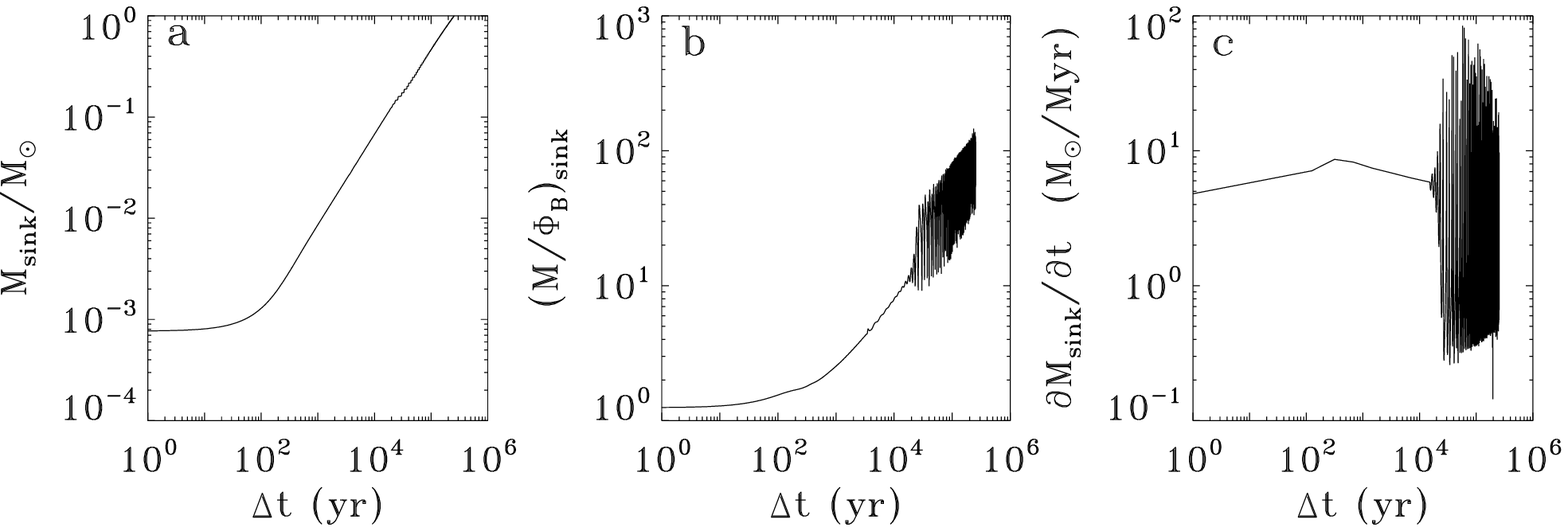}
\caption{\label{hole_quants}
(a) Time evolution of mass of the central sink (in ${\rm M_\odot}$) (b), 
Mass-to-flux ratio of central sink normalized to the critical 
value for collapse, $1/(2 \pi G^{1/2})$. (c)
Rate of mass accretion in the central sink (in ${\rm M_\odot / Myr}$).}
\end{figure}

\subsection{Time Evolution of Physical Quantities in the Forming Star}

Figure \ref{hole_quants} describes the evolution of the mass and
magnetic flux accumulated in the central sink (the forming star). The total mass 
(in ${\rm M_\odot}$) is shown in 
Figure \ref{hole_quants}a. In our model, the time required for the 
sink mass to become equal to $1 {\rm \, M_\odot}$ is $2.55\times
10^5 {\rm \,\, yr}$. For comparison, Ciolek \& K\"{o}nigl (1998) find
that in their model the corresponding time interval is only
$1.53\times 10^5 {\rm \,\, yr}$. The multiple shocks formed during the
magnetic cycles in our case, cause the mass accretion process to be
both slower and spasmodic.

The effect of the magnetic cycle on the mass accretion rate (in ${\rm M_\odot/Myr}$) is 
demonstrated directly in Figure \ref{hole_quants}c.
It initially increases because of the
ever-increasing gravitational field strength due to accumulation of
mass in the central sink. However, after the formation of the first shock 
(at $\approx 200 {\rm \,\, yr}$), the mass accretion rate decreases again,
in agreement with the findings of  Ciolek \& K\"{o}nigl (1998). 
At still later times, nevertheless, when the magnetic cycle becomes fully
developed, the mass accretion rate starts to exhibit
oscillations of large amplitude (of about one order of magnitude above
and below the mean). This can be understood in terms of the evolution
of the density in the innermost parts of the disk during one magnetic
cycle. As an outflow in the neutrals is established from a certain
radius outward, no new material accretes inward to replenish the
matter lost to the sink, and the mass accretion rate decreases
dramatically. Only after the magnetic wall is dispersed and the
outflow in the neutrals is suppressed does the density of the 
innermost part of the disk increase again, and with it the mass accretion
rate by the central sink. {\em Thus, mass accretion onto the forming star occurs in
magnetically controlled bursts}. We refer to this process as {\em spasmodic accretion}.

The mass-to-flux ratio in the sink, normalized to the critical value
for collapse ($1/2\pi\sqrt{G}$), 
also increases monotonically up to the point where the magnetic cycle
is established (Fig. \ref{hole_quants}b). From that point on, the mass-to-flux ratio exhibits
strong oscillations. Overall the mass-to-flux ratio increases by two
orders of magnitude above the critical value. This flux loss, which is
entirely due to ambipolar diffusion before the complete decoupling of
the magnetic field from the matter, is an important contribution
toward the resolution of the magnetic flux problem of star formation,
although not yet sufficient to resolve the
entire problem.

\subsection{Properties of the Shocks in the Neutrals}

%%%%%%%%%%%%%%%%%%%% FIGURE %%%%%%%%%%%%%%%%%%%%%%%
\begin{figure}
\plotone{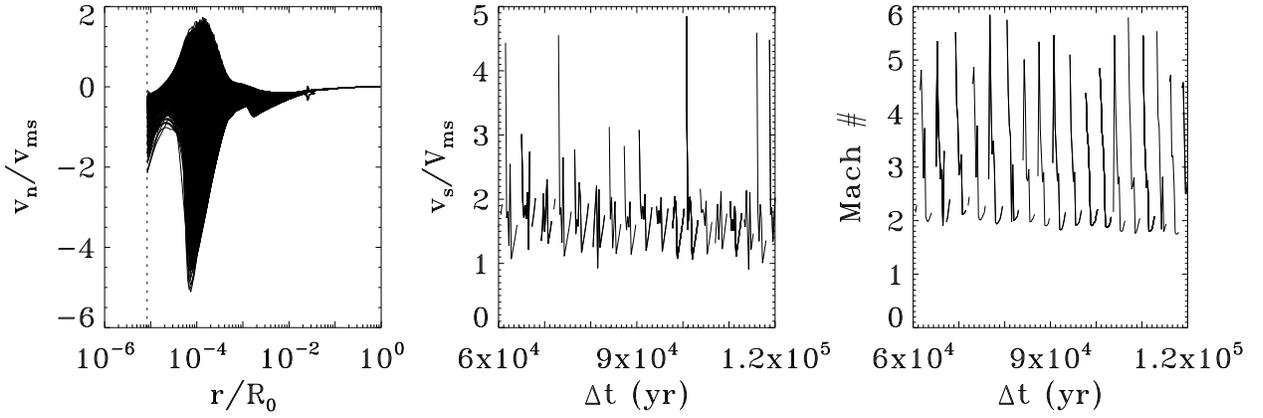}
\caption{\label{shock_quants}
(a) Velocity of the neutral fluid, normalized to the local magnetosonic
  speed $V_{\rm ms}$, as a function of $r/R_0$ at regular intervals of
  250 yr, between $\Delta t = 6 \times 10^4$ and $\Delta t = 2 \times
  10^5$ yr (b) Velocity of the shock
  normalized to $V_{\rm ms}$ as a function of time. (c)
  Mach number of shock with respect to the local magnetosonic speed 
  as a function of time.}
\end{figure}

Figure \ref{shock_quants} shows some properties of the shocks established
in the neutrals during each magnetic cycle. Figure \ref{shock_quants}a exhibits
the radial profile of the velocity of the neutral fluid normalized to
the magnetosonic speed in intervals of 250 yr for the duration of the
run. The average positions of the maximum infall velocity and the
maximum outflow velocity are near each other and remain fixed in time
during successive magnetic cycles. The magnetic cycle affects the
entire supercritical core. It is also interesting to note that although
the absolute value of the maximum infall velocity of the neutrals
increases in time (as the mass of the sink and with it the
gravitational field increase), the value of the maximum outflow
velocity remains constant with respect to the magnetosonic speed, 
and almost twice as large.

Figure \ref{shock_quants}b shows the speed of outward propagation of the shock
normalized to the magnetosonic speed,
for a series of magnetic cycles taking place between $6 \times 10^4$
and $2 \times 10^5$ yr after the introduction of the central
sink. The speed of the shock varies between 5 and 1 times the magnetosonic
speed, and generally tends to decrease as the shock propagates
outward. At almost all times the shock speed is
supermagnetosonic. This is in contrast to the results of Ciolek \&
K\"{o}nigl (1998), who found that the shock propagation is subsonic. 

Figure \ref{shock_quants}c shows the Mach number $\mathcal{M}$ with respect to the
local magnetosonic speed 
for the same time interval as in Figure \ref{shock_quants}b. In each
cycle, the shock Mach number varies between about 6 and 2, with a trend
to decrease as the shock propagates outwards. 

\subsection{Evolution of the Period of the Magnetic Cycle}

%%%%%%%%%%%%%%%%%%%% FIGURE %%%%%%%%%%%%%%%%%%%%%%%
\begin{figure}
\plotone{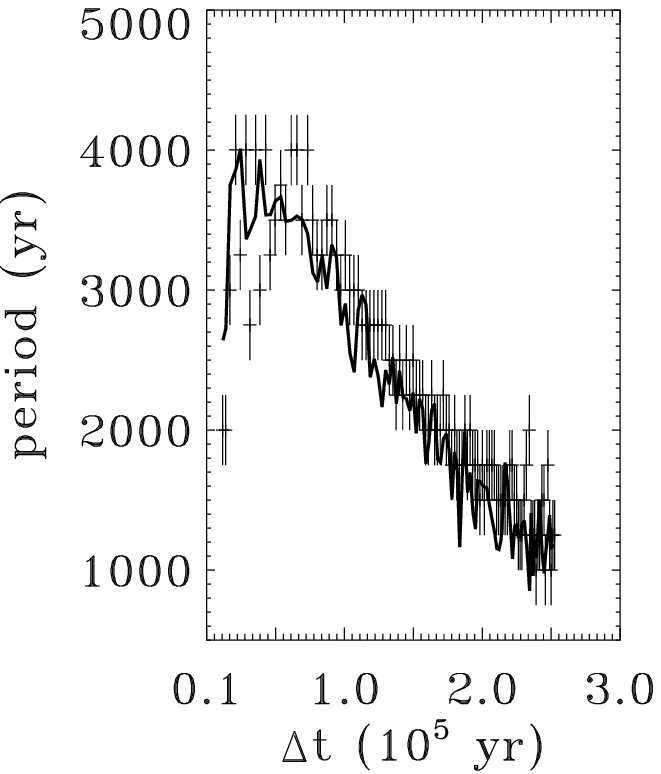}
\caption{\label{cycle_period}
Evolution in time of the period of the magnetic cycle in years (data
points). The error bars are due to the finite time resolution with
which output is obtained from the simulation. The solid curve is the 
average ambipolar-diffusion (magnetic flux-loss) timescale at the position
of the maximum infall/outflow velocity of the neutrals.}
\end{figure}

The data points in Figure \ref{cycle_period} show the period of the 
magnetic cycle in years. The ``error bars'' are due to the finite time 
resolution with which the output used to determine the duration of
each cycle is obtained from the simulation. The solid curve is the 
average ambipolar-diffusion (magnetic flux-loss) timescale,
$\tau_{\Phi}= r/2v_D$
(Mouschovias 1991, eq. [31]) at the position
of the maximum infall/outflow velocity of the neutrals. This was calculated
by averaging the drift velocity of the electrons over each cycle at the
position where the maximum neutral infall/outflow occurs. This position is,
as we already discussed, the same for all cycles. This average drift
velocity was then used to obtain the average magnetic flux-loss
timescale. {\em The agreement between the flux-loss timescale and the duration of each cycle is
excellent}. This is due to the fact that the average flux-loss timescale
represents the time required to rebuild the magnetic wall after it
is dispersed by the outflow of the electrons.
The rebuilding of the magnetic wall is the slowest phase
(bottleneck) of the cycle, and for this reason determines the entire 
duration of each cycle. 
The decrease of the duration of the magnetic cycle with time is due to
the decrease of the associated ambipolar-diffusion timescale (which
decreases as the gravitational field increases, and the maximum
infall velocity of the neutrals increases as well).  

\subsection{Stability of the Supercritical Core Against Magnetic Interchange}

\begin{figure}
\plotone{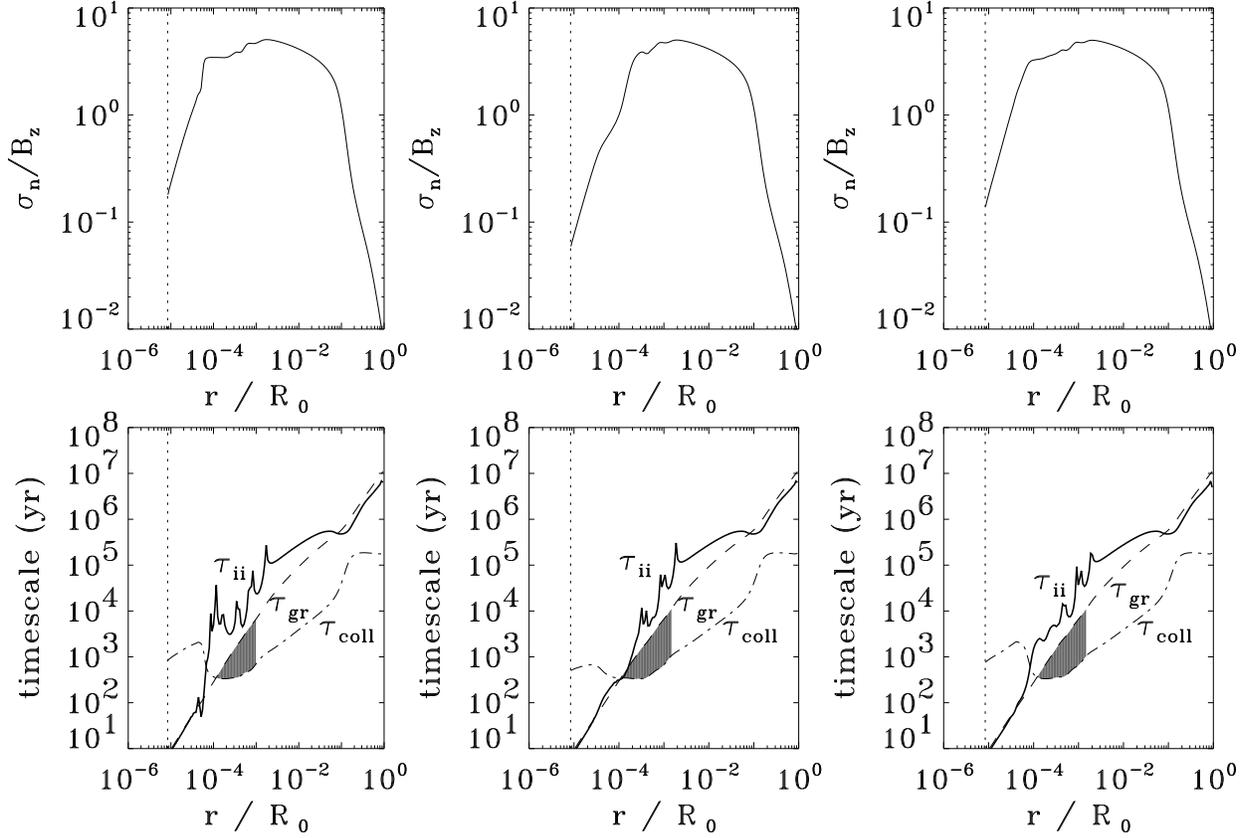}
\caption{\label{instab}
Upper panel: Local mass-to-flux ratio normalized to the 
critical value for collapse at $t=33,500$ yr (left);
$t=36,000$ yr (middle); $t=37,251$ yr (right).  
Lower panel: Gravitational contraction timescale ($\tau_{\rm gr}$,
dashed line), magnetic field - neutral fluid
collisional coupling timescale ($\tau_{\rm coll}$, dot-dashed line), 
and growthtime of most unstable mode of interchange instability 
in hydrostatic equilibrium ($\tau_{\rm ii}$, bold solid line), as
functions of $r$, at the same times as in the upper panel. The
shaded area correspond to span of allowed instability growthtimes.}
\end{figure}

Figure \ref{instab} (top row) exhibits the local mass-to-flux ratio, $dM/d\Phi_B = 
\sigma_n/B_z$ (normalized to the critical value for collapse) as a
function of position at $t=33500, 36000, 37251$ yr (from left to right) after the
introduction of the central sink (curves 1 of Fig. \ref{csafigna}; and curves
6, 8 of Fig. \ref{csafignb}, respectively). At radii smaller than the position
of the shock ($\sim 10^{-3}R_0$), the local mass-to-flux ratio
increases with increasing radius. This is a necessary
(although not sufficient) condition for the development of the
magnetic interchange instability (Spruit \& Taam 1990;  Lubow \&
Spruit 1995). That the region behind a hydromagnetic shock
in a collapsing disk is prone to the interchange instability was
speculated by Li \& McKee (1996) and also found in  
the simulation of Ciolek \& K\"{o}nigl (1998). It is therefore of
interest to examine the stability of the supercritical core with
respect to this instability. 

Two effects can prevent the magnetic interchange instability from developing:
damping due to inefficient coupling between the magnetic field and the
neutral fluid, and erasing of the perturbations due to the fast inflow
of the neutrals (Ciolek \& K\"{o}nigl 1998). 
Therefore, the instability will be able to grow only if: 
(i) its growthtime is longer that the timescale of 
collisional coupling between the magnetic field and the neutral
fluid (due to electron-neutral and/or ion-neutral collisions), 
$\tau_{\rm coll}$, and
(ii) if its growthtime is smaller than the gravitational contraction 
timescale, $\tau_{\rm gr} \approx (r / |g|)^{1/2} $. This is the 
timescale that characterizes the (dynamic) flow in this region of the supercritical core.

%%%%%%%%%%%%%%%%%%%% FIGURE %%%%%%%%%%%%%%%%%%%%%%%

In the lower panel of Figure \ref{instab}
we have plotted the gravitational-contraction (dashed line) and the collisional-coupling
(dot-dashed line)
timescales as functions of radius (normalized to $R_0$, as usual) at
the same times as in the upper panel. 
Note that in our case, the timescale of
collisional coupling between the magnetic field and the neutral fluid
is given by
\beq \label{colltime}
\frac{1}{\tau_{\rm coll}} = \frac{1}{\tau_{ne}}+\frac{\Delta_i}{\tau_{ni}},
\eeq
properly accounting for the partial attachment (or full attachment, at
certain regions in space) of the ions to the magnetic field 
(e.g., see Ciolek \& Mouschovias 1993). This is in contrast to Ciolek \&
K\"{o}nigl (1998) who used $\tau_{\rm coll} = \tau_{ni}$ (a much
smaller value than that given by eq. [\ref{colltime}]) for their
corresponding stability check, since in their model the ions were
assumed to be always attached to the magnetic field.

The shaded region in each graph in the lower panel of Figure
\ref{instab} corresponds to the span of
allowed instability growthtimes, i.e., 
growthtimes which are {\em not excluded} from developing
due to negative slope in the local mass-to-flux ratio, or due to
inefficient coupling between magnetic field and neutrals, or due to
erasing of perturbations by the neutral flow. 

The bold solid line is the growthrate for the most unstable linear
interchange mode calculated for discs in hydrostatic equilibrium
(Spruit \& Taam 1990; Lubow \& Spruit 1995), 
\beq
\tau_{\rm ii} = \left( \frac{B_{z,{\rm eq}} B_{r,Z}}{2\pi\sigma_{\rm n}}
\frac{d}{dr}\ln \frac{\sigma_{\rm n}}{B_{z,{\rm eq}}}
\right)
\,.
\eeq
Since our system is not in equilibrium, and there is no exact force
balance between magnetic and gravitational forces, the actual
instability growthtime will be {\em longer}; thus the bold line represents 
a {\em lower limit} for the instability timescale. 

The innermost region ($\lesssim 10^{-4} \approx 90$ AU), where
the slope of the local mass-to-flux ratio is steepest and the growth
of the instability is fast (small growthtime), is entirely
excluded at all times due to inefficient coupling, 
since in this region the ions are detached from the magnetic field and the
electrons are not as effective in coupling the neutrals to the field
lines. At radii greater than the point of intersection of the two curves $\tau_{\rm
  coll}$ and $\tau_{\rm gr }$, the lower limit of the instability
timescale, $\tau_{\rm ii}$, is always greater than the
contraction timescale, so perturbations are swept away before they
have an opportunity to grow. Therefore, the magnetic interchange
instability is not at all relevant.

\section{Conclusions}
 
We have studied the structure of the magnetic accretion disk
surrounding a forming star, using the central-sink
approximation to make the problem tractable numerically. 
The differences in the formulation of the problem compared to previous work 
(Ciolek \& K\"{o}nigl 1998) have been discussed in detail in Paper I.

Multiple shocks develop and propagate in the
neutral fluid, associated with a quasi-periodic magnetic cycle. 
Flux accumulation close to the central sink leads to the formation of a 
region of enhanced magnetic field strength, a {\em ``magnetic wall''}, where 
magnetic forces exceed the gravitational forces. The magnetic wall drives outflows
in the neutrals and associated shocks.
In time, the magnetic wall disperses, and the neutrals are 
reaccelerated inward behind the shock (at smaller radii), dragging with them the electrons and, 
consequently, magnetic flux (since the field is frozen in the electrons).
Flux can then reaccumulate, and the magnetic wall gets recreated, 
which causes the magnetic cycle to repeat.

The shocks in the neutrals propagate outward with supermagnetosonic
speeds in the rest frame of the central sink. 
This is in contrast to the slow accretion shock expected by Li \&
McKee (1996) and found numerically by Ciolek \& K\"{o}nigl (1998)
because they assumed that the magnetic flux is frozen in the ion fluid.
The infall velocities of the neutrals in the rest frame 
{\em of the shock} vary between 2 and 6 times the magnetosonic speed. 

The development and propagation of multiple outflows and shocks result
in mass accretion onto the forming star that occurs in {\em magnetically
controlled bursts}. We refer to this phenomenon as {\em spasmodic accretion}.
During the outflow of the neutrals, the inner part of the disk is almost ``emptied''
of matter (its column density decreases significantly) and the mass
accretion rate drops almost by two orders of magnitude below its maximum
value. It is only when the neutrals get reaccelerated inward by the
mass of the forming star that the inner
part of the disk is refilled with matter and the mass accretion rate
increases again. The time required for 1 ${\rm M_\odot}$ to 
accumulate in the
central sink is $2.55 \times 10^5 \, {\rm yr}$, characteristic of the relatively
inefficient accretion process due to the magnetically-driven, repeated outflows in the neutrals.

The radial extent of the region which is affected
by the magnetic cycle (in which the velocity of the neutrals exhibits
the quasi-periodic oscillation characteristic of the magnetic cycle)
includes almost the entire magnetically supercritical core. The magnetically
supported envelope, as expected, is not affected by the magnetic flux
redistribution in the core. 

The period of the magnetic cycle decreases in time, and is well
represented by the local flux-loss (ambipolar-diffusion) timescale. This is
expected because the bottleneck in the chain of events comprising the
magnetic cycle is the rebuilding of the magnetic wall behind the
outward propagating shock in the neutrals. This reaccumulation of
flux does indeed occur on the ambipolar-diffusion timescale, and it 
becomes faster as time progresses, since the gravitational field
(dominated by the central point mass) monotonically 
increases with time, thereby increasing the maximum infall velocity
of the neutrals (which in turn drag the
electrons inward). 

We found that the magnetically supercritical core is stable against interchange-type 
instabilities. This result is in contrast to the predictions 
by Li \& McKee (1996) and Ciolek \& K\"{onigl} (1998). The reason for
this difference with the results of Ciolek \& K\"{o}nigl 
is the fact that they assumed flux-freezing in
the ions, which is not valid above $\sim 10^8$ ${\rm cm^{-3}}$ and which
leads to much better coupling between the neutrals and the magnetic
field in their calculation. 

Ambipolar diffusion by itself during the accretion phase significantly contributes
to the resolution of the magnetic flux problem of star formation by 
increasing the mass-to-flux ratio of the protostar by more than two
orders of magnitude above its critical value. However, this is not
sufficient to resolve the entire magnetic flux problem yet. 

\acknowledgements{ We thank Glenn Ciolek, Chester Eng and Vasiliki Pavlidou
for useful discussions. This work was carried out without external support, 
and would not have been published without the generosity of the ApJ, and the
Astronomy and Physics Departments of the University of Illinois. KT was 
supported by the University of Illinois Research Board and the Greek State 
Scholarship Foundation during this project.}

%%%%%%%%%%%%%%%%%%%%%%%%%%%%%%%%%%%%%%%%%%%%%%%%%%%%%%%
% end of main text
%%%%%%%%%%%%%%%%%%%%%%%%%%%%%%%%%%%%%%%%%%%%%%%%%%%%%%%

\end{document}